# The Persistent Buffer Tree:
# An I/O-efficient Index for Temporal Data

Saju J. Dominic [1], G. Sajith[1]

[1] Indian Institute of Technology, Guwahati, India.
{sjdom,sajith}@iitg.ernet.in

**Abstract.** In a variety of applications, we need to keep track of the development of a data set over time. For maintaining and querying this multi version data I/O-efficiently, external memory data structures are required. In this paper, we present a probabilistic self-balancing persistent data structure in external memory called the persistent buffer tree, which supports insertions, updates and deletions of data items at the present version and range queries for any version, past or present. The persistent buffer tree is I/O-optimal in the sense that the expected amortized I/O performance bounds are asymptotically the same as the deterministic amortized bounds of the (single version) buffer tree in the worst case. ...

## 1 Introduction

The study of I/O-efficient algorithms and data structures has been receiving increased attention in recent years due to the fact that communication between fast internal memory and slower external memory such as disks is the bottleneck in many computations involving massive datasets. The significance of this bottleneck is increasing as internal computation gets faster and especially as parallel computing gains popularity.

The importance of maintaining data not only in their latest version, but also to keep track of their development over time has been widely recognized. Version data in engineering databases and time-oriented data are two prime examples. A lot of work has already been done in developing persistent data structures in external memory [4, 5], but all of these data structures are designed to be used in on-line settings, where queries should be answered immediately and within a good worst case number of I/O's. This effectively means that using these structures to solve offline problems yields non-optimal algorithms because they are not able to take full advantage of the large internal memory.

In this paper, we present an I/O-optimal probabilistic self-balancing persistent data structure called the *persistent buffer tree* that supports operations such as insertions, updates, deletions and range queries within an expected optimal number of I/O's in the worst case. This is achieved by using random priorities to balance the search tree and by only requiring good amortized performance of the operations on the structure, and by allowing search operations to be batched. The persistent buffer tree is *partially persistent* in the sense that updates can only be applied to the present version whereas queries can be applied on any version, present or past.

## 1.1 Model of computation

We will be working in an I/O model introduced by Aggarwal and Vitter [1]. Our computational model consists of a single processor with a small internal memory connected to a large external memory. The model has the following parameters:

- $N$ = the number of elements in the problem instance;
- $M$ = the number of elements that can fit into internal memory;
- $B$ = the number of elements per block;
- $Z$ = the number of elements in the output,

where $M < N$ and $1 << B < M/2$. The model captures the essential parameters of many of the I/O-systems in use today, and depending on the size of the data elements, typical values are of the order of $M = 10^6$ or $10^7$ and $B = 10^3$. Large scale problem instances can be in the range $N = 10^{10}$ or $10^{12}$.

An I/O operation in the model is a swap of $B$ elements from internal memory with $B$ consecutive elements from external memory. The measure of performance we consider is the number of such I/O's needed to solve a given problem. The quotients N/B (the number of blocks in the problem) and M/B (the number of blocks that fit into internal memory) and Z/B (the number of blocks in the output) play an important role in the study of I/O-complexity. Therefore we will use $n$ as shorthand for N/B, $m$ for M/B and $z$ for Z/B.

## 1.2 Previous Results

Aggarwal and Vitter [1] first considered the problem of designing I/O-efficient algorithms. They gave several algorithms for basic problems such as sorting, permuting, and matrix operations. However, until recently, research in I/O-efficient algorithms centered on the fundamental problems of sorting, permuting, and the like. Goodrich et al. [7] were the first to develop external memory techniques that would apply to wider classes of algorithms — in their case, computational geometry. Later, Chiang et al. [8] opened up the area of external-memory graph algorithms. In most of this work, the data structures used by the algorithms were motivated by the particular problems that were being considered, rather than being the focus of the work. Arge [2] improved several of the results in these earlier papers by the introduction of the I/O efficient *buffer tree*, which was the first I/O-efficient data structure to incorporate an amortized analysis for batched operations. Arge's buffer tree was developed to assist in the design of algorithms in computational geometry, supporting operations such as range searching. However, it also led to simple algorithms for sorting and some graph problems, generalizing some of the results of Chiang et al. . Kumar et al. [3] introduced the I/O-efficient heap and tournament tree based on the buffer tree. Becker et al. [4] and Varman and Verma [5] developed partially persistent versions of B-trees that support insertion, updates and deletes in ? (log$_B$N) I/O's per operation. Salzberg and Tsotras [6] survey work done on persistent access methods and other techniques for time-evolving data.

### 1.3 Our results

In this paper, we present an I/O-optimal partially persistent randomized data structure called the *persistent buffer tree* that supports insert, update and delete in expected optimal $O\left(\frac{\log_m n}{B}\right)$ I/O's amortized and range search operation in expected optimal $O\left(\frac{\log_m n}{B} + z\right)$ I/O's amortized. We improve the bounds for insertion, update and deletion from ? (log$_B$N) to $O\left(\frac{\log_m n}{B}\right)$, roughly a factor of O(B log$_B$m) improvement over the previous best known result. We achieve this bound by introducing persistency to a randomized version of Arge's buffer tree at no increase in the asymptotic I/O performance bounds.

In particular, the persistent buffer tree supports the following operations:

- insert(key, info) : inserts a new element with the given key and info, into the present version, at an optimal expected amortized cost of $O\left(\frac{\log_m n}{B}\right)$ I/O's; this operation creates a new version.

- update(key, info) : updates the info of the element with the given key, into the present version, at an optimal expected amortized cost of $O\left(\frac{\log_m n}{B}\right)$ I/O's; this operation creates a new version.

- delete(key) : deletes the (unique) element with the given key from the present version , at an optimal expected amortized cost of $O\left(\frac{\log_m n}{B}\right)$ I/O's; this operation creates a new version.

- range search(lowkey, highkey, version) : return all elements whose key lies between the given lowkey and the given highkey in the given version, at an optimal expected amortized cost of $O\left(\frac{\log_m n}{B} + z\right)$ I/O's; this operation does not create a new version.

As mentioned some work has already been done on designing persistent data structures in external memory, but all of it has been done in an I/O model where the size of the internal memory equals the block size. The motivation for working in this model has been partly been that the goal was to develop structures for an online setting, where answers to the queries should be reported immediately and within a good worst case number of I/O's. For typical systems $B$ is less than $M$ so log$_B$N is larger than log$_m$n, but more important, the persistent B-tree solution will be slower than the optimal solution by a factor of $B$. As $B$ typically is on the order of thousands this factor is crucial in practice. The main problem with the persistent B-tree in this context is precisely that it is designed to have a good worst case online search performance. In order to take advantage of the large internal memory, we on the other hand, use the fact that we are only interested in the *overall* I/O use of the algorithm for an *offline* problem - that is, in a good amortized performance of the operations – and even satisfied with batched search operations.

The remainder of the paper is organized as follows. In Section 2, we present our persistent buffer tree and establish amortized bounds on its performance. In Section 3, we discuss our conclusions.

## 2 The Persistent Buffer Tree

In this Section, we present our technique to transform a single version buffer tree into a multi version probabilistic self-balancing buffer tree with the same I/O performance bounds.

### 2.1 The Basic Idea

To achieve the desired behavior, we associate insertion and deletion versions with elements, since elements of different life spans need to be stored in the same block. We follow the convention that each update (insert or delete) operation creates a new version; the *i*-th update creates version *i*. The special version value $ is used to indicate the present version. An element is considered to be *live* if it has been neither updated nor deleted, and *dead* otherwise. An element inserted by update operation *i* into the tree carries a lifespan of [i,$) at the time of insertion; deletion of an element by update operation *i* changes its deletion version from $ to *i*.

We also associate a random priority with each element. The random priority is chosen independently for each element from a continuous uniform distribution. The random priority is used to balance the structure as the tree is arranged in heap order with respect to priorities and in-order with respect to key values.

The persistent buffer tree is a *m*–ary tree, with each node having a buffer of size M, holding upto m blocks of elements. Each internal node holds ?(m) blocks of elements and the tree has an expected height of $O(\log_m n)$ provided the value of m is large[x].

The operations on the structure, updates as well as queries, are done in a lazy manner. If we, for example, want to insert an element in the tree, we generate a random priority, but we do not, as in persistent B-tree, search down the tree to find the place among the leaves to insert the element. Instead, we wait until we have collected a block of insertions (and/or other operations), and then we insert this block in the buffer of the root node. When the buffer becomes full, we apply the updates and push all but the largest m/2 blocks of elements with respect to the priority values, one level down the tree, in search order of the key values, to the buffers on the next level. We call this the buffer-emptying process. Updates, as well as queries, are basically done in the same way as insertions. It means that queries get batched in the sense that the result of a query may be generated (and reported) lazily by several buffer-emptying processes.

The main requirement needed to satisfy the I/O bounds mentioned in the introduction is that we should be able to empty a buffer in a number of I/O's that is linear in the number of elements in the buffer. If this is the case, we can do an amortization argument by associating a number of credits to each block of elements in the tree. More precisely, each block in the list or buffer of node x must hold O(the height of the tree rooted at x) credits. As we only empty a buffer when it becomes full, the blocks in the buffer can pay for the emptying process as they all get pushed one level down. We give each update element $O(\frac{\log_m n}{B})$ credits and each query element $O(\frac{\log_m n}{B} + z)$ credits on insertion in the root node, and this gives us the desired bounds.

## 2.2 Description of the data structure

The persistent buffer tree is a random m-way search tree with the following properties:

- Each element is denoted by *<key, in_version, del_version, priority, info>*. The *key* value is unique for any given version and the lifespan of the item is from its insertion version *in_version* to its deletion version *del_version*. The priority is a random value generated independently for each element from a continuous uniform distribution.
- Each node has *m* children.
- Each node has a *buffer* of size M associated with it, which contains up to *m* blocks of elements, stored in sorted order w.r.t. priority value.
- Any element held in a node's *buffer* has a smaller priority value than any other element stored in the buffers of its child nodes.
- Any element held in a node's *buffer* has no smaller key value than any other element stored in the buffers of its left siblings and has no larger *key* value than any other element stored in the buffers of its right sibling.
- Every internal node of the tree holds ?(m) blocks of elements.
- The tree defines a *m*-ary heap over priority values and a m-ary search tree over the key values.

We shall now state an important result on the expected height of a random m-way search tree, due to Devroye[9].

**Lemma 1:** The expected height of a random m-way search tree over n elements is O(log$_m$ n).

*Proof*: A random m-ary search tree is constructed from a random permutation of 1, 2,...,n. A law of large numbers is obtained for the height H$_n$ of these trees by applying the theory of branching random walks. In particular, it is shown that H$_n$/log n ? ? in probability as n ? 8. where ? = ?(m) is a constant depending upon m only. Interestingly, as m? 8, ?(m) is asymptotic to 1/(log m), the coefficient of *log* n in the asymptotic expression for the height of the complete m-ary search tree. This proves that for large m, random m-ary search trees behave virtually like complete m-ary trees. See[9] for the detailed proof.

## 2.3  The Persistent Buffer Tree Primitives

The standard operations *insert*, *update and delete* are supported, and are implemented using the EMPTYBUFFER primitive.

### 2.3.1  EMPTYBUFFER

EMPTYBUFFER is executed whenever a node's buffer has over *m* blocks of elements. This primitive is used to push elements down one level in the tree in search order of the key values, so that the heap order is satisfied with respect to the priority values. The steps involved in EMPTYBUFFER are:

1. The *m* blocks of elements are loaded into main memory in O(m) I/O's. The elements are then sorted w.r.t. the key value as primary key and in_version as secondary key.

   Step 1 requires ?(m) I/O's as m blocks of elements are loaded into main memory.

2. In case there are elements with identical key values, perform updates and deletes on each such group as described below( Note that a zero in_version denotes a delete as e xplained in Section 2.4) .

   2.1. [Handling update] The del_version of each item (having non zero in_version) is set to the in_version of the next item.

   2.2. [Handling delete] The del_version of the last item is set to del_version of the item with zero in_version (if such an item exists).

   Step 2 requires no I/O's as all computations are done on elements in main memory.

3. The elements in the buffer are sorted w.r.t priority values and all but the smallest m/2 blocks of elements(with respect to priority) are distributed to the children's buffer in search order of the key values. The distribution is done in such a way that the number of elements in each child buffer is as evenly balanced as possible . Step 4 requires ?(m) I/O's as ?(m) blocks of elements are pushed down one level of the tree to O(m) child buffers .

4. If any of the children's buffer now contains more than *m* blocks of elements, EMPTYBUFFER is recursively executed on the child buffer.

We will now state with proof s ome simple lemmas about EMPTYBUFFER.

**Lemma 2.1**:  The cost of EMPTYBUFFER is ?(m)  I/O's.
*Proof*: Step 1 requires ?(m) I/O's as *m* blocks of elements are loaded into main memory. Step 2 requires no I/O's as all computations are done on elements in main memory. Step 3 requires ?(m) I/O's as ?(m) blocks of elements are pushed down one level of the tree to O(m) child buffers. The result follows.

**Lemma 2.2**: EMPTYBUFFER maintains the invariant that each internal node holds ?(m) blocks of elements.

*Proof*: The number of elements in the buffer after Step 1 and Step 2 is *m* blocks. Step 3 is executed after which the number of elements in the buffer is *m/2* blocks. Thus at the end of EMPTYBUFFER, the buffer has between m/2 elements. The result follows .

**Lemma 2.3**: The next EMPTYBUFFER on any node is guaranteed not to occur for the next O(vm) blocks of updates (or other operations).

*Proof*: This is a simple consequence of Lemma 2.2 . Since at the end of EMPTYBUFFER, the buffer has *m/2* blocks of elements and EMPTYBUFFER is executed only when a node's buffer has over *m* blocks of elements, the next EMPTYBUFFER on the buffer does not occur for the next *m/2* updates(or other operations). The result follows .

**Lemma 2.4**: EMPTYBUFFER maintains the invariant that any element held in a node's buffer has no smaller key value than any other element held in the lists of any of its child nodes.

*Proof*: Assume that the invariant is true at the beginning of Step 1 of EMPTYBUFFER. Steps 1 and 2 do not affect the invariant property. Step 3 maintains the invariant as the larger elements, with respect to priority, are retained in the buffer and lighter elements, with respect to priority, are pushed one level down the tree.. Hence if the invariant holds at the beginning of EMPTYBUFFER, then it also holds at the end of EMPTYBUFFER. The result follows .

**Lemma 2.5**: EMPTYBUFFER maintains the invariant that any element held in a node's buffer has no smaller key value than any other element stored in the lists of its left siblings and no greater key value than any other element stored in the lists of its right sibling.

*Proof*: This is guaranteed by the manner in which the elements in the buffer are distributed to the buffers of the children(in search order of the key values) in Step 3 of EMPTYBUFFER. The other Steps do not affect the invariant property. The result follows .

### 2.4 The Persistent Buffer Tree Operations

We are now ready to describe the operations supported by the persistent buffer tree, namely insert, update, delete and range query.

#### 2.4.1 insert (key, info )

The insert operation is used to insert a new element with the given *key*, containing the given *info* , into the tree at the current version.

The steps involved in *insert* are:

1. Construct a new element consisting of the new element to be inserted, having the given key and info.

2. The in_version of the element is set to the current version number. The current
   version number is incremented by one. The del_version is set to $.
3. When $B$ such insert elements have been collected in internal memory, insert the
   block in the buffer of the root node.
4. If the buffer of the root node now contains more than $m$ blocks of such elements,
   perform a EMPTYBUFFER on the node as explained in Section 2.3.1 .

We will now state with proof, a simple theorem about insert operation.

**Theorem 1.1:** A new element is inserted in to the tree at an expected amortized cost of
$O\left(\frac{\log_m n}{B}\right)$ I/O's which is optimal.

*Proof*: The expected height of the tree is $O(\log_m n)$ by Lemma 1. The EMPTYBUFFER
primitive moves $?(m)$ blocks of insert elements one level down the tree in an amortized
cost of $?(m)$ I/O's by Lemma 2.1 or one insert element is moved down one level of the
tree in $O(1/B)$ I/O's amortized. The lower bound of $O(n \log_m n)$ on sorting [1] together
with the bound on tree height implies that the expected amortized cost of $O\left(\frac{\log_m n}{B}\right)$
per insertion is optimal. The result follows.

### 2.4.2 update (key, info)

The update operation is used to update an existing element with the given *key*, with
the given *info*, at the current version.

The steps involved in *update* are:
1. Construct a new (update) element with the given key and the new info.
2. The in_version of the element is set to the current version number. The current
   version number is incremented by one. The del_version is set to $.
3. When $B$ such update elements have been collected in internal memory, insert the
   block in the buffer of the root node.
4. If the list of the root node now contains more than $m$ blocks of such elements,
   perform a EMPTYBUFFER on the node as explained in Section 2.3.1 .

We will now state with proof, a simple theorem about update operation.

**Theorem 1.2:** An existing element in the tree is updated at an expected amortized cost
of $O\left(\frac{\log_m n}{B}\right)$ I/O's which is optimal.

*Proof*: The expected height of the tree is $O(\log_m n)$ by Lemma 1. The EMPTYBUFFER
primitive moves $?(m)$ blocks of update elements one level down the tree in an amor-
tized cost of $?(m)$ I/O's by Lemma 2.1 or one update element is moved down one level
of the tree in $O(1/B)$ I/O's amortized. The lower bound of $O(n \log_m n)$ on sorting [1]
together with the bound on tree height implies that the expected amortized cost of
$O\left(\frac{\log_m n}{B}\right)$ I/O's per update is optimal. The result follows.

### 2.4.3 delete (key)

The delete operation is used to delete an existing element with the given *key*, at the current version. An element once deleted from the persistent database is not allowed to be inserted or updated subsequently.

The steps involved in *delete* are:

1. Construct a new (delete) element with the given key.

2. The in_version of the element is set to zero. The del_version is set to the current version number. The current version number is incremented by one.

3. When *B* such delete elements have been collected in internal memory, insert the block in the buffer of the root node.

4. If the buffer of the root node now contains more than *m* blocks of such elements, perform a EMPTYBUFFER on the node as explained in Section 2.3.1 .

We will now state with proof, a simple theorem about delete operation.

**Theorem 1.3:** An existing element in the tree is deleted at an amortized cost of $O\left(\frac{\log_m n}{B}\right)$ I/O's which is optimal.

*Proof:* The expected height of the tree is $O(\log_m n)$ by Lemma 1. The EMPTYBUFFER primitive moves $?(m)$ blocks of delete elements one level down the tree in an amortized cost of $?(m)$ I/O's by Lemma 2.1 or one delete element is moved down one level of the tree in $O(1/B)$ I/O's amortized. The lower bound of $O(n \log_m n)$ on sorting [1] together with the bound on tree height implies that a cost of $O\left(\frac{\log_m n}{B}\right)$ per update would be optimal. The result follows.

### 2.4.3 range search (lowkey, highkey, version)

The range search operation is used to report all elements having a key value between given *lowkey* and given *highkey* that belongs to the given version. An element is said to belong to version *i* if its life span contains *i* .

The general idea in our range search operation is as follows: We start almost as when we do an insertion or a deletion. We make a new element containing the interval [lowkey, highkey] and the given version and insert it into the root buffer. We then have to modify our buffer-emptying process in order to deal with the new range search elements. The basic idea is that when we meet a range search element in a buffer-emptying process, we first determine whether lowkey and highkey are contained in the same subtree among the subtrees rooted at the children of the node in question. If this is the case, we just insert the element in the corresponding buffer. Otherwise, we split the element into two, one for lowkey and one for highkey, and report the elements in the subtrees for which the elements are in the given interval and version. The splitting occurs only once and after that range search elements are pushed downwards in the buffer-emptying processes like insert and delete elements, while elements in the sub-trees which satisfy the search criteria are reported.

We will now state with proof, a simple theorem about range search operation.

**Theorem 1.4:** An existing element in the tree is updated at an optimal amortized cost of $O(\frac{\log_m n}{B} + r)$ I/O's, where is $r$ is the number of blocks of reported elements.

*Proof*: The expected height of the tree is $O(\log_m n)$ by Lemma 1. A range search element is moved down one level of the tree in $O(1/B)$ I/O's amortized by Lemma 2.1 and 3.1, while the elements in the subtree satisfying the search criteria are reported in $r$ I/O's. The result follows.

### 2.5 Complexity Analysis of the Persistent Buffer Tree

### 2.5.1 I/O Complexity of Persistent Buffer Tree

The primitive `EMPTYBUFFER` performs all the I/O operations, moving a collection of elements one level down or up the tree. This functions are I/O efficient — that is, when it moves x elements up or down one level, it need $O(x)$ I/O's to do so.

We are now ready to state with proof our main theorem.

**Theorem 2:** *The total expected cost of an arbitrary sequence of N intermixed insert, update, delete and range search operations performed on an initially empty persistent buffer tree is $O(n\log_m n + r)$ I/O operations. Here r.B is the number of reported elements.*

*Proof*: The persistent buffer tree performs insert, update and delete operations on an element in an optimal expected amortized $O(\frac{\log_m n}{B})$ I/O's by Theorem 1.1. , 1.2 and 1.3 respectively. The range search operation is performed at an optimal expected amortized $O(\frac{\log_m n}{B} + r)$ I/O's by Theorem 1.4. The result follows.

### 2.5.2 Space Complexity of Persistent Buffer Tree

The number of internal nodes in the tree is *n/m* and each internal node has a buffer of size M. Each element is stored at only one location in the tree. Hence the space complexity of the tree is $O(n)$.

## 3. Conclusion

In this paper, we have developed a technique for transforming a single version buffer tree into a multi version buffer tree at no increase in the asymptotic I/O complexity. The persistent (multi version) buffer tree has the same expected I/O performance bounds as the (single version) buffer tree for the operations of insert, update, and delete and range search in the worst case.